\documentstyle[aps,prd,epsf]{revtex}

\newcommand{\be}{\begin{equation}}
\newcommand{\ee}{\end{equation}}
\newcommand{\ben}{\begin{eqnarray}
\displaystyle}
\newcommand{\een}{\end{eqnarray}}

\newcommand{\la}{{\lambda}}

\newcommand{\de}{{\delta}}

\newcommand{\cJ}{{\cal J}}

\newcommand{\p}{\partial}
\newcommand{\na}{\nabla}
\newcommand{\tna}{{\tilde \nabla}}
\newcommand{\hna}{{\hat \nabla}}

\newcommand{\Liek}{{\cal L}_{k^{\mu}}}

\newcommand{\tg}{{\tilde g}}

\newcommand{\hg}{{\hat g}}

\newcommand{\hR}{{\hat R}}
\newcommand{\hSi}{{\hat \Sigma}}

\newcommand{\trho}{{\tilde \rho}}
\newcommand{\th}{{\tilde h}}

\newcommand{\tG}{{\tilde G}}
\newcommand{\tH}{{\tilde H}}
\newcommand{\tR}{{\tilde R}}

\newcommand{\ep}{\epsilon}

\newcommand{\ga}{\gamma}

\begin{document}

\title{Uniqueness Theorem for Generalized Maxwell Electric and Magnetic Black
Holes
in Higher Dimensions}

\author{Marek Rogatko}

\address{Institute of Physics \protect \\
Maria Curie-Sklodowska University \protect \\
20-031 Lublin, pl.Marii Curie-Sklodowskiej 1, Poland \protect \\
rogat@tytan.umcs.lublin.pl \protect \\
rogat@kft.umcs.lublin.pl}

\date{\today}

\maketitle
\begin{abstract}
Based on the conformal energy theorem 
we prove the uniqueness theorem for static higher dimensional {\it electrically} and
{\it magnetically} charged black holes being the solution
of Einstein $(n-2)$-gauge forms equations of motion.
Black holes
spacetime 
contains an asymptotically flat spacelike hypersurface with compact interior and
non-degenerate components of the event horizon.
\end{abstract}


\baselineskip=18pt
\section{Introduction}
The problem of classification of non-singular black hole solutions was first raised by
Israel \cite{isr},
M\"uller zum Hagen {\it et al.} \cite{mil73} and Robinson \cite{rob77},
while
the most complete results were proposed in Refs.\cite{bun87,ru,ma1,he1,he93}.
The classification of static vacuum black hole solutions was finished 
in \cite{chr99a}, where
the condition of non-degeneracy of the 
event horizon was removed. In Ref.\cite{chr99b} the Einstein-Maxwell (EM) black holes
were treated and it was shown that for the static
electro-vacuum black holes all degenerate components of 
the event horizon should have charges of the same signs.
\par
The problem of the uniqueness black hole theorem for stationary axisymmetric
spacetime was considered in Refs.\cite{stat}. But the complete proof was delivered by
Mazur \cite{maz} and Bunting \cite{bun}
(see for a review of the uniqueness of black hole
solutions story see \cite{book} and references therein).
\par
Due to the attempts of building a consistent quantum gravity theory there was also
resurgence of works concerning the mathematical aspects of the low-energy string theory
black holes.
The staticity theorem for Einstein-Maxweel axion dilaton (EMAD) gravity
was studied in Ref.\cite{sta}. Then,
the uniqueness of the black hole solutions in dilaton gravity was  proved in 
works
\cite{dil,mar01}, while the uniqueness of the static
dilaton $U(1)^2$
black holes being the solution of $N = 4, d = 4$ supergravity
was provided in \cite{rog99}. The extension of the proof to the theory
to allow for the inclusion of 
$U(1)^{N}$ static dilaton black holes was established in Ref.\cite{rog02}.
\par
The recent unification attempts such as M/string theory
reveal the concept that our Universe
may be a brane or defect emerged in higher dimensional geometry. 
$E8 \times E8$ heterotic string theory
at strong coupling may be described in terms of M-theory acting in eleven-dimensional
spacetime with boundaries where ten-dimensional Yang-Mills gauge theories reside on two
boundaries \cite{hor}. One hopes, that this idea will be helpful
in solving hierarchy problem.
All these
trigger the interests in higher
dimensional black hole solutions. 
The so-called
TeV gravity attracts attention to higher dimensional black hole 
which may be produced in high energy experiments \cite{gid}. 
\par
The considerable interests was also attributed to
$n$-dimensional black hole uniqueness theorem, both in vacuum and
charged case \cite{gib03,gib02a,gib02b,kod04}. The complete 
classification of $n$-dimensional
charged black holes having both degenerate and non-degenerate
components of event horizon was provided in Ref.\cite{rog03}.
Proving the uniqueness theorem for stationary 
$n$-dimensional black holes is much more complicated.
It turned out that generalization of Kerr metric to arbitrary $n$-dimensions proposed by
Myers-Perry \cite{mye86} is not unique. The counterexample showing that a five-dimensional
rotating black hole ring solution with the same angular momentum and mass
but the horizon of which was homeomorphic to $S^{2} \times S^{1}$ was presented in
\cite{emp02} (see also Ref.\cite{emp04}). Recently, it has been established that
Myers-Perry solution is the unique black hole in five-dimensions in the class of 
spherical topology
and three commuting Killing vectors \cite{mor04}.
\par
The uniqueness theorem for self-gravitating nonlinear $\sigma$-models
in higher dimensional spacetime was obtained in \cite{rog02a}.
The $n$-dimensional black hole uniqueness theorems 
for supersymmetric black holes in five-dimensions
were given in Refs.
\cite{sup}.
\par
In Ref.\cite{mye86} it was pointed out that a black hole being the source of both 
magnetic and electric components of $2$-form $F_{\mu \nu}$ was a striking coincidence.
In order to treat
this problem in $n$-dimensional gravity we shall consider
both {\it electric} and {\it magnetic} components of 
$(n-2)$-gauge form $F_{\mu_{1} \dots \mu_{n-2}}$. We shall comprise
the uniqueness of {\it electrically} and {\it magnetically} charged static
$n$-dimensional black hole solution containing an asymptotically
flat hypersurface with compact interior with non-degenerate components
of the event horizon. The main result of our paper will be the proof of the uniqueness
of static higher dimensional {\it electrically} and {\it magnetically} charged
black hole containing an asymptotically flat hypersurface with compact interior
and non-degenerate components of the event horizon.

\section{Higher dimensional generalized Einstein-Maxwell system}
In this section we shall examine the generalized Maxwell $(n-2)$-gauge
form $F_{\mu_{1} \dots \mu_{n-2}}$ in $n$-dimensional spacetime
described by the following
action:
\be
I = \int d^n x \sqrt{-{g}} \bigg[ {}^{(n)}R - 
F_{(n-2)}^2
\bigg],
\label{act}
\ee
where ${g}_{\mu \nu}$ is $n$-dimensional metric tensor,
$F_{(n-2)} = dA_{(n-3)} $ is $(n-2)$-gauge form field.
The metric of $n$-dimensional static spacetime subject to the
asymptotically timelike Killing vector field 
$k_{\alpha} = \big({\p \over \p t }\big)_{\alpha}$
and $V^{2} = - k_{\mu}k^{\mu}$ can be written in the following form:
\be
ds^2 = - V^2 dt^2 + g_{i j}dx^{i}dx^{j},
\label{met}
\ee
where $V$ and $g_{i j}$
are independent of the $t$-coordinate as the quantities
of the hypersurface $\Sigma$ of constant $t$. \\
The energy momentum tensor of the $(n-2)$-gauge form 
$T_{\mu \nu} = - {\de S \over \sqrt{- g}\de g^{\mu \nu}}$ yields
\be
T_{\mu \nu} = (n -2) F_{\mu i_{2} \dots i_{n-2}}
F_{\nu}{}{}^{i_{2} \dots i_{n-2}} - {g_{\mu \nu} \over 2}F_{(n-2)}^2.
\ee
\\
In our consideration we shall take into account the asymptotically
flat spacetime, i.e., the spacetime will contain a data set 
$(\Sigma_{end}, g_{ij}, K_{ij})$ with gauge fields of $F_{(n-2)}$
such that $\Sigma_{end}$ is diffeomorphic to $\bf R^{n-1}$ minus 
a ball. The asymptotical conditions of the following forms should also be satisfied:
\ben
\mid g_{ij} - \delta_{ij} \mid +~ r \mid \p_{a} g_{ij} \mid + \dots
+r^{m} \mid \p_{a_{1} \dots a_{m}} g_{ij} \mid + r \mid K_{ij}\mid + \dots
+r^{m} \mid \p_{a_{1} \dots a_{m}} K_{ij} \mid \le {\cal O}\bigg( {1 \over r} \bigg), \\ 
\mid F_{i_{1} \dots i_{n-2}} \mid +~ 
r \mid \p_{a} F_{i_{1} \dots i_{n-2}} \mid + \dots +
r^{m} \mid \p_{a_{1} \dots a_{m}}F_{i_{1} \dots i_{n-2}} \mid
\le {\cal O}\bigg( {1 \over r^{2}} \bigg).
\een
We define {\it electric} $(n-3)$-form by the following expression:
\be
E_{i_{1} \dots i_{n-3}} = F_{i_{1} \dots i_{n-2}} k^{i_{n-2}},
\ee
and {\it magnetic} $1$-form as
\be
B_{k} = {1 \over \sqrt{2 (n -2)!}}~ \ep_{k \mu i_{1} \dots i_{n-2}}
F^{i_{1} \dots i_{n-2}} k^{\mu}.
\ee
We introduce also the rotation $(n-3)$-form of the stationary
Killing vector field $k_{\mu}$
\be
\omega_{j_{2} \dots j_{n-2}} = {(n-2) \over \sqrt{2 (n -2)!}}~
\ep_{j_{2} \dots j_{n-2} \mu \nu \gamma} k^{\mu} \na^{\nu} k^{\gamma}.
\label{tw}
\ee
Directly from the definition (\ref{tw}) and definition of {\it electric} and
{\it magnetic} forms and equations of motion for $(n-2)$-gauge form
we find Eqs. of motion for {\it magnetic} $1$-form $B_{k}$
\be
\na^{a}\bigg( {B_{a} \over N }\bigg) = {E^{i_{2} \dots i_{n-2}}~
\omega_{i_{2} \dots i_{n-2}} \over N^2},
\label{bbb}
\ee
and similarly for {\it electric} $(n-3)$-form $E_{i_{1} \dots i_{n-3}}$
\be
\na^{j_{1}} \bigg( {E_{j_{1} j_{2}  \dots j_{n-3}}\over N}\bigg) = - {B^{a}~
\omega_{a j_{2} \dots j_{n-2}}\over N^2},
\label{eee}
\ee
where we have denoted $N = k_{\mu}k^{\mu}$.
\par
In an asymptotically flat, globally hyperbolic spacetime with compact bifurcation surface
and strictly stationary, simply connected domain of outer communication
$<<\cJ>>$ (i.e., $V^2 = - k_{\mu}k^{\mu} \ge 0$ ) one has that $<<\cJ>>$
is static if
\be
\omega_{j_{2} \dots j_{n-2}} = 0 \qquad
\Leftrightarrow \qquad k^{\alpha} R_{\alpha [\beta }k_{\ga ]} = 0.
\ee
This statement can be verified by the direct use of Eqs. of motion
and by using the stationarity conditions for $F_{\mu_{1} \mu_{2} \dots \mu_{n-2}}$
with respect to the Killing vector field $k_{\mu}$, namely
\be
\Liek F_{\mu_{1} \dots \mu_{n-2}} = 0.
\ee
Then, by the direct calculation one can check that
$R_{\alpha \beta}k^{\beta} = 0$ and this provides the statement.
In our paper we deal with the static charged black hole case which
implies that
in static  domain of outer communication
$<<\cJ>>$ the right-hand sides of Eqs.(\ref{bbb}) and (\ref{eee}) are equal to zero.
Having in mind the exact form of the metric tensor (\ref{met}) 
on the hypersurface $\Sigma$ orthogonal to the Killing vector field $k_{\mu}$
and assuming
the only one non-trivial {\it electric}
component of the $(n-2)$-form $F_{\mu_{1} \dots
\mu_{n-2}}$ as $A_{0 1 \dots n-2} = \phi(x)$ and for the
{\it magnetic} potential $B_{k} = {}^{(g)}\na_{k} g(x)$,
we get the following equations of motion:
\be
{}^{(g)}\na_{i}{}^{(g)}\na^{i} V =
{(n - 3)^2 \over V}{}^{(g)}\na_{i} \phi {}^{(g)}\na^{i} \phi
+ {2 (n - 3)\over (n - 2) V} {}^{(g)}\na_{i} g {}^{(g)}\na^{i}g,
\ee
\be
{}^{(g)}\na_{i} {}^{(g)}\na^{i} \phi = {1 \over V}{}^{(g)}\na_{i} \phi
{}^{(g)}\na^{i} V,
\ee
\be
{}^{(g)}\na_{i} {}^{(g)}\na^{i} g = {1 \over V}{}^{(g)}\na_{i} g
{}^{(g)}\na^{i} V,
\ee
\ben
{}^{(n-1)} R_{ij} &=& {{}^{(g)}\na_{i}{}^{(g)}\na_{j}V \over V}
- {n - 2 \over V^2} {}^{(g)}\na_{i}\phi {}^{(g)}\na_{j} \phi
+ {g_{ij} \over V^2}{}^{(g)}\na_{i}\phi {}^{(g)}\na^{i} \phi \\ \nonumber
&-& {2 \over V^2} {}^{(g)}\na_{i}g {}^{(g)}\na_{j}g + {2 g_{ij} \over (n - 2) V^2}
{}^{(g)}\na_{i} g {}^{(g)}\na^{i} g.
\een
The covariant derivative 
with respect to 
$g_{ij}$ is denoted by ${}^{(g)}\na$,
while ${}^{(n - 1)}R_{ij}(g)$ is the Ricci tensor defined on 
the hypersurface $\Sigma$.
\par
Let us assume further
that in asymptotically flat spacetime there is a standard coordinates
system in which we have the usual asymptotic expansion
\be
V = 1 - {\mu \over r^{n - 3}} + {\cal O}\bigg( {1 \over  r^{n - 2}} \bigg),
\ee
and for the metric tensor
\be
g_{ij} = \bigg( 1 + {2 \over n - 3}{\mu \over r^{n - 3}} \bigg)+
{\cal O} \bigg( {1 \over r^{n - 2}} \bigg).
\ee
While for {\it electric} and {\it magnetic} potential one gets respectively
\ben
\phi = {Q/C_{1} \over r^{n - 3}} + {\cal O} \bigg( {1 \over r^{n - 2}} \bigg), \\
g = {P/C_{2} \over r^{n - 3}} + {\cal O} \bigg( {1 \over r^{n - 2}} \bigg),
\een
where $C_{1} = n - 3$, ${C_{2}} = 2(n-3)/n-2$,
$\mu$ is the ADM mass seen by the observer from the infinity, 
$Q$ is the electric charge, $P$ is magnetic charge while
$r^2 = x_{i}x^{i}$.
Let us define the quantities in the forms as follows:
\ben
\Phi_{1} &=& {1 \over 2} \bigg[ V + {1 \over V} - {(n - 2) \phi^2 \over V}
\bigg], \\
\Phi_{0} &=& {\sqrt{n - 2}~ \phi \over V},\\
\Phi_{-1} &=& {1 \over 2} \bigg[ V - {1 \over V} - {(n - 2) \phi^2 \over V}
\bigg],
\een
and
\ben
\Psi_{1} &=& {1 \over 2} \bigg[ V + {1 \over V} - {2 (n - 3) g^2 \over V}
\bigg], \\
\Psi_{0} &=& {\sqrt{2 (n - 3)}~ g \over V},\\
\Psi_{-1} &=& {1 \over 2} \bigg[ V - {1 \over V} - {2 (n - 3) g^2 \over V}
\bigg].
\een
Furthermore,
if we define the metric $\eta_{AB} = diag(1, -1, -1)$, one can check that
\be
\Phi_{A} \Phi^{A} = \Psi_{A} \Psi^{A} = -1.
\ee
Next consider the following conformal transformation:
\be
\tg_{ij} = V^{2 \over n - 3} g_{ij},
\ee
and 
introduce the symmetric tensors written as
\be 
\tG_{ij} = \tna_{i} \Phi_{-1} \tna_{j} \Phi_{-1} -
\tna_{i} \Phi_{0} \tna_{j} \Phi_{0} -
\tna_{i} \Phi_{1} \tna_{j} \Phi_{1},
\label{ggg}
\ee
and similarly for the potential $\Psi_{A}$
\be
\tH_{ij} = \tna_{i} \Psi_{-1} \tna_{j} \Psi_{-1} -
\tna_{i} \Psi_{0} \tna_{j} \Psi_{0} -
\tna_{i} \Psi_{1} \tna_{j} \Psi_{1},
\label{hhh}
\ee
where $\tna_{i}$ is the covariant derivative with respect to the metric $\tg_{ij}$.
By virtue of relations (\ref{ggg}) and (\ref{hhh}) the field equations
imply
\be
\tna^{2}\Phi_{A} = \tG_{i}{}{}^{i} \Phi_{A}, \qquad
\tna^{2} \Psi_{A} = \tH_{i}{}{}^{i} \Psi_{A},
\label{ppff}
\ee
where $A = - 1, 0, 1$, 
and it is straightforward to establish that
the Ricci tensor of the metric $\tg_{ij}$ can be expressed as
\be
\tR_{ij} = \tG_{ij} + {1 \over n - 3}\tH_{ij}.
\label{rr}
\ee
In the next step we study the conformal transformations given by the expressions
\be
{}^{\Phi}g_{ij}^{\pm} = {}^{\phi}\omega_{\pm}^{{2 \over n - 3}} \tg_{ij},
\qquad
{}^{\Psi}g_{ij}^{\pm} = {}^{\psi}\omega_{\pm}^{{2 \over n - 3}} \tg_{ij},
\ee
where the conformal factors are determined by
\be
{}^{\Phi}\omega_{\pm} = {\Phi_{1} \pm 1 \over 2}, \qquad
{}^{\Psi}\omega_{\pm} = {\Psi_{1} \pm 1 \over 2}.
\label{pf}
\ee
Just one gets four manifolds $(\Sigma_{+}^{\Phi}, {}^{\Phi}g_{ij}^{+})$,
$(\Sigma_{-}^{\Phi}, {}^{\Phi}g_{ij}^{-})$, $(\Sigma_{+}^{\Psi}, {}^{\Psi}g_{ij}^{+})$,
$(\Sigma_{-}^{\Psi}, {}^{\Psi}g_{ij}^{+})$. Pasting 
$(\Sigma_{\pm}^{\Phi}, {}^{\Phi}g_{ij}^{\pm})$ and 
$(\Sigma_{\pm}^{\Psi}, {}^{\Psi}g_{ij}^{\pm})$ across the surface $V = 0$
we can construct regular hypersurfaces $\Sigma^{\Phi} = \Sigma_{+}^{\Phi}
\cup \Sigma_{-}^{\Phi}$
and $\Sigma^{\Psi} = \Sigma_{+}^{\Psi}
\cup \Sigma_{-}^{\Psi} $. 
If $(\Sigma, g_{ij}, \Phi_{A}, \Psi_{A})$ are asymptotically flat solution of
(\ref{ppff}) and (\ref{rr}) with non-degenerate black hole event horizon,
our next task will be to check that total
gravitational mass on hypersurfaces $\Sigma^{\Phi}$ and $\Sigma^{\Psi}$ 
is equal to zero. In order to do this we shall implement the conformal
positive mass theorem in higher dimensions \cite{gib02a,sim99}. Now using
another conformal transformation given by
\be
\hg^{\pm}_{ij} = \bigg[ \bigg( {}^{\Phi}\omega_{\pm} \bigg)^2
 \bigg( {}^{\Psi}\omega_{\pm} \bigg)^{2 \la} \bigg]^{{1 \over (n-3)(1 + \la)}}\tg_{ij},
\ee
it follows that 
the Ricci curvature tensor on the space yields
\ben \label{ric}
(1 + \la) \hR &=& \bigg[ {}^{\Phi}\omega_{\pm}^2 {}^{\Psi}\omega_{\pm}^{2 \la} \bigg]
^{-1 \over (n - 3)(1 + \la)}
\bigg( {}^{\Phi}\omega_{\pm}^{2 \over n - 3} {}^{\Phi}R +
\la {}^{\Psi}\omega_{\pm}^{2 \over n - 3} {}^{\Psi}R \bigg) \\ \nonumber
&+& {\la \over 1 + \la} \bigg( {n - 2 \over n - 3} \bigg)
\bigg( \hna _{i} \ln {}^{\Phi}\omega_{\pm} - \hna _{i} \ln {}^{\Psi}\omega_{\pm} \bigg)  
\bigg( \hna ^{i} \ln {}^{\Phi}\omega_{\pm} - \hna ^{i} \ln {}^{\Psi}\omega_{\pm} \bigg).  
\een
For this stage on we shall take $\la = 1 /n-3$.\\
The close inspection of
the first term in Eq.(\ref{ric}) reveals that it is non-negative. Namely
one can establish that it may be written in the form as follows:
\ben
{}^{\Phi}\omega_{\pm}^{2 \over n - 3} {}^{\Phi}R +
\la {}^{\Psi}\omega_{\pm}^{2 \over n - 3} {}^{\Psi}R &=& 
\bigg( {n - 2 \over n - 3} \bigg) \mid
{\Phi_{0} \tna_{i} \Phi_{-1}
- \Phi_{-1} \tna_{i} \Phi_{0} \over
\Phi_{1} \pm 1 } \mid^2 \\ \nonumber
&+&
{(n - 2) \over (n - 3)^2} \mid { \Psi_{0} \tna_{i} \Psi_{-1}
- \Psi_{-1} \tna_{i} \Psi_{0} \over
\Psi_{1} \pm 1} \mid^2.
\een

Applying the conformal energy theorem we draw a conclusion that
$(\Sigma^{\Phi}, {}^{\Phi}g_{ij})$, $(\Sigma^{\Psi}, {}^{\Psi}g_{ij})$ and
$(\hSi, \hg_{ij})$ are flat and it in turns implies that the conformal factors
${}^{\Phi}\omega = {}^{\Psi}\omega$ and $\Phi_{1} = \Psi_{1}$. Furthermore
$\Phi_{0} = const~ \Phi_{-1}$ and $\Psi_{0} = const~ \Psi_{-1}$. Just the above potentials 
are functions of a single variable. Moreover, the manifold $(\Sigma, g_{ij})$ is conformally flat.
We can rewrite $\hg_{ij}$ in a 
conformally flat form \cite{gib03,gib02a}, i.e., we define a function
\be
\hg_{ij} = {\cal U}^{4 \over n-3} {}^{\Phi}g_{ij},
\label{gg}
\ee
where ${\cal U} = ({}^{\Phi}\omega_{\pm} V)^{-1/2}$.
Just, it turned out that $\hR$ is equal to zero provided
the Einstein $(n-2)$-gauge form equations of motion reduced
to the Laplace equation on the $(n - 1)$ Euclidean manifold
\be
\na_{i}\na^{i}{\cal U} = 0,
\ee
where $\na$ is the connection on a flat manifold. 
Having in mind the above Eq. we can imply
the following expression for the flat base space
\be
{}^{\Phi}g_{ij} dx^{i}dx^{j} = \trho^{2} d{\cal U}^2 + \th_{AB}dx^{A}dx^{B},
\ee
The event horizon is located at some ${\cal U} = const$ and
one can show that the embedding of $\cal H$ into the Euclidean
$(n-1)$ space is totally umbilical \cite{kob69}. 
Each of the connected components of the horizon $\cal H$ will be
a geometric sphere of a certain radius.
The radius can be
determined by the value of
$\rho \mid_{\cal H}$,
where $\rho$ is the coordinate which can be introduced on $\Sigma$
as follows:
$$\hg_{ij}dx^{i}dx^{j} = \rho^2 dV^2 + h_{AB}dx^{A}dx^{B}.$$
Thus, it is clearly seen that the embedding is hyperspherical.\\
Of course one can always locate one 
connected component of the horizon at $r = r_{0}$ surface without loss of generality.
Thus we have a boundary value problem for the Laplace equation 
on the base space $\Omega = E^{n-1}/B^{n-1}$ with the
Dirichlet boundary condition
(the rigid embedding \cite{kob69}).
Let ${\cal U}_{1}$ and ${\cal U}_{2}$ be two solutions of the boundary value problem.
By successive use of
the Green identity and integrating over the volume element
we find
\be
\bigg( \int_{r \rightarrow \infty} - \int_{\cal H} \bigg) 
\bigg( {\cal U}_{1} - {\cal U}_{2} \bigg) {\p \over \p r}
\bigg( {\cal U}_{1} - {\cal U}_{2} \bigg) dS = \int_{\Omega}
\mid \na \bigg( {\cal U}_{1} - {\cal U}_{2} \bigg) \mid^{2} d\Omega.
\ee
The surface integrals vanish due to the imposed boundary conditions provided that
the volume integral must be identically equal to zero.
\par
Hence the preceding results can be collected in the following.\\
{\it Theorem}:\\
Consider a static solution to $n$-dimensional Einstein $(n-2)$-gauge forms
$F_{\mu_{1} \dots \mu_{n-2}}$ equation of motion with only {\it electric}
and {\it magnetic} charge. Let us suppose that we have an asymptotically
timelike Killing vector field $k_{\mu}$ orthogonal to the connected and simply
connected spacelike hypersurface $\Sigma$. The topological
boundary $\p \Sigma$ of $\Sigma$ is a nonempty topological manifold with
$g_{ij}k^{i} k^{j} = 0$ on $\p \Sigma$. Thus, we obtain the following conclusion:\\
If $\p \Sigma$  is connected, then there exist a neighbourhood of the
hypersurface $\Sigma$ which is diffeomorphic to an open
set of a generalized Reissner-Nordsr\"om non-extreme solution with
{\it electric} and {\it magnetic} charges provided by the adequate     
{\it electric} and {\it magnetic} components of
the gauge $(n-2)$-form $F_{\mu_{1} \dots \mu_{n-2}}$.

\vspace{0.5cm}



\end{document}